\documentclass
[floats,amsmath,amssymb,prb,letterpaper,showpacs,onecolumn]{revtex4}%
\pdfoutput=1
\usepackage{graphicx}
\usepackage{amsmath}
\usepackage{amsfonts}
\usepackage{amssymb}%
\providecommand{\U}[1]{\protect\rule{.1in}{.1in}}
\ifx\pdfoutput\relax\let\pdfoutput=\undefined\fi
\newcount\msipdfoutput
\ifx\pdfoutput\undefined\else
\ifcase\pdfoutput\else
\ifx\paperwidth\undefined\else
\ifdim\paperheight=0pt\relax\else\pdfpageheight\paperheight\fi
\ifdim\paperwidth=0pt\relax\else\pdfpagewidth\paperwidth\fi
\fi\fi\fi
\begin{document}
\title{Nanoscale Magnetic Heat Pumps and Engines }
\author{Gerrit E. W. Bauer and Stefan Bretzel}
\affiliation{Kavli Institute of NanoScience, Delft University of Technology, Lorentzweg 1,
2628 CJ Delft, The Netherlands}
\author{Arne Brataas}
\affiliation{Department of Physics, Norwegian University of Science and Technology, NO-7491
Trondheim, Norway}
\author{Yaroslav Tserkovnyak}
\affiliation{Department of Physics and Astronomy, University of California, Los Angeles,
California 90095, USA}
\date{\today}

\begin{abstract}
We present the linear response matrix for a sliding domain wall in a rotatable
magnetic nanowire, which is driven out of equilibrium by temperature and
voltage bias, mechanical torque, and magnetic field. An expression for
heat-current induced domain wall motion is derived. Application of Onsager's
reciprocity relation leads to a unified description of the Barnett and
Einstein-de Haas effects as well as spin-dependent thermoelectric properties.
We envisage various heat pumps and engines,\textit{ }such as coolers driven by
magnetic fields or mechanical rotation as well as nanoscale motors that
convert temperature gradients into useful work. All parameters (with the
exception of mechanical friction) can be computed microscopically by the
scattering theory of transport.

\end{abstract}

\maketitle

\section{Introduction}

Onsager's reciprocal relations\cite{Onsager} reveal that seemingly unrelated
phenomena can be expressions of identical microscopic correlations between
thermodynamic variables of a given system.\cite{Groot} The archetypal example
is the Onsager-Kelvin identity of thermopower and Peltier cooling.

Research into the interaction between electric currents and the ferromagnetic
order parameter of the last years has paid off handsomely. On one hand, the
predicted charge current-induced spin transfer torque\cite{Berger,Slonczewski}
in metallic ferromagnetic structures such as spin valves and domain
walls\cite{Tatara,Klaui,Ralph} in ferromagnetic wires has been understood in
some detail and applied to random access magnetic memories,\cite{Hitachi}
logics,\cite{Matsunaga} and shift registers.\cite{Parkin} On the other hand,
it has been established that a moving magnetization pumps a spin
current\cite{Tserkovnyak02} that can be converted into charge currents and
voltages in ferromagnet$|$normal metal bilayers,\cite{Wang} ferromagnetic
textures,\cite{Barnes} multilayers,\cite{Xiao} or by the spin Hall
effect.\cite{Saitoh} Spin-pumping induced voltages have been observed in
metallic magnetic heterostructures,\cite{Saitoh,Costache} tunnel
junctions\cite{Moriyama} and magnetic wires with a moving domain
wall.\cite{Yang} A modern illustration of the power of Onsager's relations is
the demonstration that spin-transfer torques and charge currents induced by
magnetization dynamics are two sides of the same
medal.\cite{Saslow,Duine,Mecklenburg}

Domain walls also react to thermal gradients, as first observed and discussed
by Jen and Berger.\cite{Berger85,Jen1,Jen2} Domain wall displacement by laser
heating is a possible technology for high-density magnetic
recording.\cite{Miyakoshi} Hatami \textit{et al}.\cite{Hatami} and
Saslow\cite{Saslow} proposed a thermoelectric spin-transfer torque as
mechanism for magnetization switching in spin valves and domain wall motion in
magnetic wires. More recently, Kovalev \textit{et al}.\cite{Kovalev09}
addressed this issue for general one-dimensional spin textures in
ferromagnetic wires. Yuan \textit{et al}. found large non-adiabatic
corrections to the thermal torques on narrow domain walls.\cite{Yuan09}

The scattering theory of electron transport can be employed to describe
dissipative processes in magnetic systems such as the Gilbert damping of
magnetizations dynamics,\cite{Brataas08,Starikov} leading to a microscopic
formalism for the Onsager coefficients that govern the interaction between
charge currents and magnetization dynamics.\cite{Hals,Hals2}

Nearly a century ago it was discovered that in macroscopic bodies the
magnetization of a ferromagnet couples to the mechanical degree of freedom:
Barnett\ demonstrated that a mechanical rotation of a demagnetized ferromagnet
creates a net magnetization along the rotation axis,\cite{Barnett} whereas
Einstein and de Haas showed that reversing the magnetic moment of a
ferromagnetic cylinder induces a mechanical torque.\cite{Einstein} Both
effects are governed by the same gyromagnetic tensor.\cite{LLP} The
microscopic theory of mechanical and magnetic angular momentum coupling in
nanostructures has recently been picked up
again.\cite{Fulde,Kovalev03,Kovalev05,Kovalev07} Moreover, Wallis \textit{et
al}.\cite{Wallis} succeeded in measuring an Einstein-de Haas effect by
agitating a magnetic cantilever. Zolfagharkhani \textit{et al}%
.\cite{Zolfagharkhani} detected the mechanical torque induced by the decay of
a current-induced magnetization, which can be interpreted as a variation of
the Einstein-de Haas effect.\cite{Kovalev08} The conditions to observe the
Barnett effect in nanostructures have been estimated by Bretzel \textit{et
al}.\cite{Bretzel}

In this paper we show that effects of magnetic, electric, thermal, and
mechanical forces can be unified in a linear response matrix relating the
conjugated thermodynamic variables for charge, energy, magnetization, and
mechanical rotation. In order to keep mathematics simple, we focus on a thin
wire of an easy-plane metallic ferromagnet as studied by Hals \textit{et
al.}\cite{Hals} for current-induced magnetization dynamics. The wire is
connects two heat and particle baths and is allowed to rotate (Figure 1). We
may profit from Onsager's relations according to which we have to fill in only
one half of the nondiagonal elements of the response matrix. This implies, for
example, that in the linear regime Barnett and Einstein-de Haas effects are
equivalent. We identify the heat-current driven domain wall motion and
conclude that domain wall motion is associated with the pumping of heat (or a
\textquotedblleft thermal motive force\textquotedblright). The mechanical
torque generated by a temperature difference opens the vista of magnetic
nanoscale heat engines.

In this paper we first recapitulate the basic thermodynamics following de
Groot\cite{Groot} for a conventional thermoelectric element in Section IIa. In
Section IIb we show that the Onsager principles can be applied to the coupling
between magnetic and mechanical dynamics for a model system of a magnetic wire
containing a domain wall. In Section III the Onsager matrix is derived for a
coupled thermoelectric and magnetomechanical system. In Section IV we specify
how the Onsager matrix elements can be computed microscopically. Section V is
devoted to a discussion of the magnitude of the couplings for a model system
consisting of a nanowire of a ferromagnetic metal wire encapsulated in
mult0-wall carbon nanotubes. Section VI summarizes the conclusions.

\section{Non-equilibrium thermodynamics}

According to the second law of thermodynamics the entropy $\mathcal{S}$ is
maximal for the equilibrium state such that
\begin{equation}
\Delta\mathcal{S}=-\frac{1}{2}\sum_{i=1}^{n}\sum_{k=1}^{n}g_{ik}a_{i}a_{k}%
\leq0
\end{equation}
for small deviations of the $n$ state variables $a_{i}=A_{i}-\bar{A}_{i}$ from
their equilibrium values $\bar{A}_{i}$. The matrix of coefficients $\hat{g}$
is positive definite and symmetric. If we parameterize a small deviation of
the system from thermodynamic equilibrium by the forces (affinities) $X_{i}$
defined by (where $T$ is the equilibrium temperature):
\begin{equation}
X_{i}\equiv T\frac{\partial\mathcal{S}}{\partial a_{i}}=-T\sum_{k=1}^{n}%
g_{ik}a_{k}\,, \label{forces}%
\end{equation}
then, in linear response, the variables $A_{i}$ will relax to their
equilibrium values $\bar{A}_{i}$ according to
\begin{equation}
J_{i}\equiv\dot{a}_{i}=\sum_{k=1}^{n}L_{ik}X_{k}, \label{response}%
\end{equation}
defining the response matrix $\hat{L}$. Its elements can be introduced
phenomenologically or computed from microscopic principles by the
Kubo-Greenwood formalism or scattering theory. The system responses $J_{i}$
are called fluxes, currents, rates, velocities \textit{etc}. Eq.
(\ref{response}) remains valid in the presence of external forces slowly
varying in time that may render the $\bar{A}_{i}$ time dependent. The entropy
generation rate reads%
\begin{equation}
\dot{\mathcal{S}}=-\sum_{i}\dot{a}_{i}\sum_{k}g_{ik}a_{k}=\frac{1}{T}\sum
_{i}J_{i}X_{i}. \label{Sgeneration}%
\end{equation}
Onsager discovered that, due to microscopic time-reversal symmetry, the linear
response coefficients obey the reciprocity relations
\begin{equation}
L_{ik}\left(  \mathbf{H}_{\mathrm{ext}},\mathbf{m}\right)  =\epsilon
_{i}\epsilon_{k}L_{ki}\left(  -\mathbf{H}_{\mathrm{ext}},-\mathbf{m}\right)  .
\label{OR}%
\end{equation}
where $\epsilon_{i}=1$ if the state variable $a_{i}$ is even under time
reversal and $\epsilon_{i}=-1$ otherwise. Time-reversal (anti)symmetry in the
presence of external magnetic fields $\mathbf{H}_{\mathrm{ext}}$ and
equilibrium magnetic ordering indicated by a vector field with unit length
$\mathbf{m}(\mathbf{r})$ (parametrizing the position-dependent direction of
the magnetization) has been made explicit. The inverse of the response matrix
$\hat{L}:$
\begin{equation}
X_{i}=\sum_{k=1}^{n}L_{ik}^{-1}J_{k},
\end{equation}
has the same Onsager symmetry%
\begin{equation}
L_{ik}^{-1}\left(  \mathbf{H}_{\mathrm{ext}},\mathbf{m}\right)  =\epsilon
_{i}\epsilon_{k}L_{ki}^{-1}\left(  -\mathbf{H}_{\mathrm{ext}},-\mathbf{m}%
\right)  . \label{ORI}%
\end{equation}

\subsection{Thermoelectric element}

Consider as an example an ordinary thermoelectric element (such as a wire)
connecting two reservoirs which are in respective thermal equilibria but at
different temperatures $T_{1}/T_{2}$ and voltages $V_{1}\ $/$V_{2}$. Let us
define $\Delta T=T_{2}-T_{1}\ll T$ and $\Delta V=V_{2}-V_{1}$. If the wire has
no independent degrees of freedom, we can describe a general (slightly
out-of-equilibrium) state of this closed system by (half of) the energy and
charge differences between the two reservoirs, $U=(U_{2}-U_{1})/2$ and
$q=(q_{2}-q_{1})/2$, respectively. Disregarding the wire's heat capacity and
electrostatic capacitance relative to those of the large reservoirs, $U$ and
$q$ correspond to the energy and charge that have been transferred from
reservoir 1 (left) to reservoir 2 (right) with respect to some reference
state. $J_{c}=\dot{q}$ and $J_{Q}=\dot{U}$ are, respectively, charge and
energy currents associated with $U$ and $q$ that are driven by $\Delta T$ and
$\Delta V$. We next employ the thermodynamic identity%
\begin{equation}
T_{j}\dot{\mathcal{S}}_{j}=\dot{U}_{j}-V_{j}\dot{q}_{j}\,,
\end{equation}
which holds for each reservoir separately. To leading order in the
perturbations, the total entropy change $\dot{\mathcal{S}}=\dot{\mathcal{S}%
}_{1}+\dot{\mathcal{S}}_{2}$ introduced by moving a small amount of energy and
charge between the reservoirs is thus
\begin{equation}
T\mathcal{\dot{S}}=-\frac{\Delta T}{T}\dot{U}-\Delta V\dot{q}%
\end{equation}
By comparison with Eq.~(\ref{Sgeneration}), we identify the conjugate fluxes
and forces:
\begin{equation}
J_{Q}=\dot{U}\,,\;X_{Q}=-\frac{\Delta T}{T}\,;\;J_{c}=\dot{q}\,,\;X_{c}%
=-\Delta V\,,
\end{equation}
such that Eq. (\ref{response}) becomes%
\begin{equation}
\left(
\begin{array}
[c]{c}%
J_{c}\\
J_{Q}%
\end{array}
\right)  =\left(
\begin{array}
[c]{cc}%
L_{11} & L_{12}\\
L_{21} & L_{22}%
\end{array}
\right)  \left(
\begin{array}
[c]{c}%
-\Delta V\\
-\frac{\Delta T}{T}%
\end{array}
\right)  .
\end{equation}
The Onsager matrix can be rewritten in terms of the electric conductance
\begin{equation}
G=-\left.  \frac{J_{c}}{\Delta V}\right\vert _{\Delta T=0},
\end{equation}
heat conductance
\begin{equation}
\kappa=-\left.  \frac{J_{Q}}{\Delta T}\right\vert _{J_{c}=0},
\end{equation}
and thermopower or Seebeck coefficient
\begin{equation}
S=\left.  -\frac{\Delta V}{\Delta T}\right\vert _{J_{c}=0},
\end{equation}
such that%
\begin{equation}
\left(
\begin{array}
[c]{c}%
J_{c}\\
J_{Q}%
\end{array}
\right)  =\left(
\begin{array}
[c]{cc}%
G & GTS\\
GTS & TG\left(  \frac{\kappa}{G}+TS^{2}\right)
\end{array}
\right)  \left(
\begin{array}
[c]{c}%
-\Delta V\\
-\frac{\Delta T}{T}%
\end{array}
\right)  . \label{JJ}%
\end{equation}
Traditionally, the role of currents and voltages in the thermoelectric
response are exchanged. In terms of the resistance $R=1/G:$%
\begin{equation}
\left(
\begin{array}
[c]{c}%
-\Delta V\\
J_{Q}%
\end{array}
\right)  =\left(
\begin{array}
[c]{cc}%
R & -TS\\
TS & T\kappa
\end{array}
\right)  \left(
\begin{array}
[c]{c}%
J_{c}\\
-\frac{\Delta T}{T}%
\end{array}
\right)  .
\end{equation}
Hereby, we recovered the Onsager-Kelvin relation between thermopower and
Peltier coefficient:
\begin{equation}
\Pi\equiv\left.  \frac{J_{Q}}{J_{c}}\right\vert _{\Delta T=0}=TS\,.
\end{equation}
The Sommerfeld approximation leads to the Wiedemann-Franz law $\kappa
=G\mathcal{L}T\ $and Mott's formula $S=-e\mathcal{L}T\partial_{E}\ln
G|_{E_{F}},$ where $\partial_{E}\ln G|_{E_{F}}$ is the logarithmic energy
derivative of the conductance at the Fermi energy $E_{F}$, $\mathcal{L}%
=(k_{B}/e)^{2}\pi^{2}/3$ is the Lorenz number and $-e$ the electron charge.
The dimensionless expression $S^{2}/\mathcal{L}$ vanishes quadratically at low
temperatures, is small for most metals at room temperature,\cite{Hatami} and
may usually be disregarded. Eq. (\ref{JJ}) then becomes
\begin{equation}
\left(
\begin{array}
[c]{c}%
J_{c}\\
J_{Q}%
\end{array}
\right)  =G\left(
\begin{array}
[c]{cc}%
1 & TS\\
TS & \mathcal{L}T^{2}%
\end{array}
\right)  \left(
\begin{array}
[c]{c}%
-\Delta V\\
-\frac{\Delta T}{T}%
\end{array}
\right)  \,, \label{TE}%
\end{equation}

\subsection{Magnetomechanical element \label{MM}}

We consider now a quasi-one-dimensional magnetic nanowire with easy-plane
anisotropy that contains a transverse domain wall, which is the standard model
system for the study of magnetic domain wall motion. We chose here the
tail-to-tail (rather than head-to-head) topology shown in Figure 1. The wire
is mounted in a low-friction bearing such that it can freely rotate around its
($x$) axis and a mechanical torque $\tau_{\mathrm{ext}}^{\mathrm{mech}}$ can
be applied. The system can also be driven by an applied magnetic field
$H_{\mathrm{ext}}$, and, via electric and thermal contacts, by a voltage
$\left(  \Delta V\right)  $ and/or temperature $\left(  \Delta T\right)  $
bias.
\begin{figure}[ptb]
\centering
\includegraphics[
natheight=3.804200in,
natwidth=10.268900in,
height=2.2923in,
width=6.1418in
]{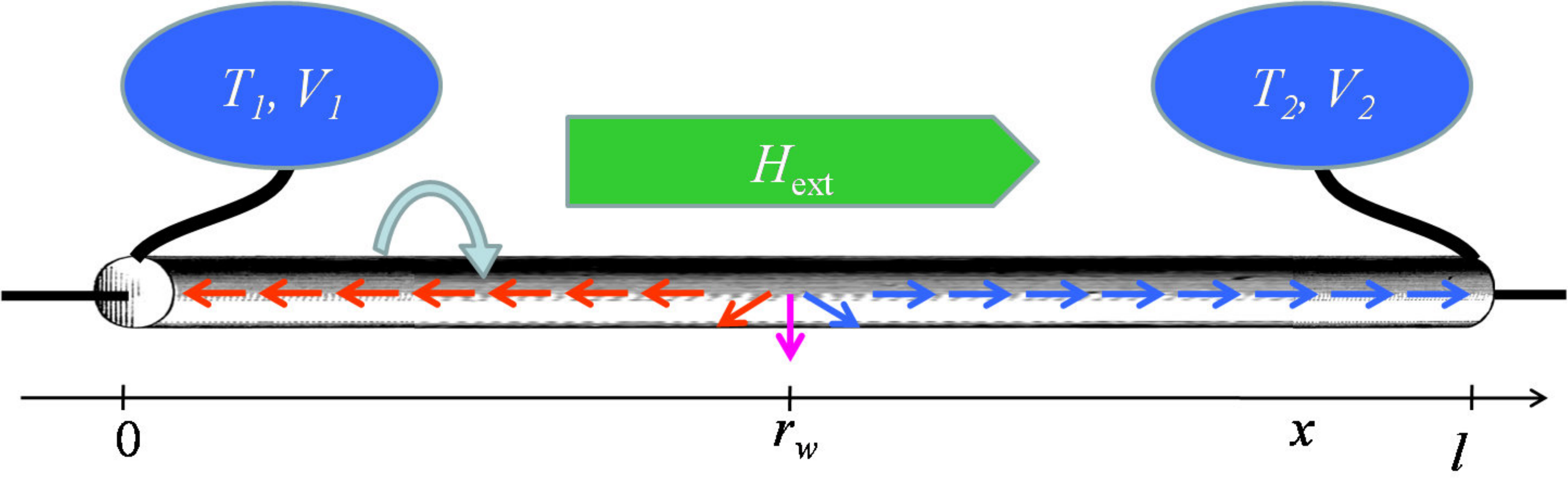}\caption{Magnetic nanowire of length $l$ in electrical and thermal
contact with reservoirs. A domain wall is centered at position $r_{w}$. The
wire is mounted such that it can rotate around the $x$-axis. A magnetic field
and mechanical torque can be applied along $x.$}%
\label{System}%
\end{figure}

Let us suppose initially that the magnetomechanical properties are decoupled
from the electric and heat currents. The equation of motion of the
magnetization $M_{s}\mathbf{m}\left(  x,t\right)  \mathbf{,}$ where $M_{s}$ is
the constant saturation magnetization, is governed by the
Landau-Lifshitz-Gilbert (LLG) equation, appended by Barnett's gauge field that
represents the aligning torque felt by angular momenta in rotating systems. In
the frame of reference that rotates with the wire:\cite{Bretzel}
\begin{equation}
\dot{\mathbf{m}}=-\gamma\mathbf{m}\times\mathbf{H}_{\mathrm{eff}}%
+\alpha\mathbf{m}\times\dot{\mathbf{m}}+\mathbf{m}\times\mathbf{x}\dot
{\varphi}\,, \label{LLG}%
\end{equation}
where $\gamma$ is the minus the gyromagnetic ratio ($\gamma>0$ for electrons)
and $\dot{\varphi}$ the angular velocity of the wire around its axis. The
effective field $\mathbf{H}_{\mathrm{eff}}$ is the functional derivative of
the free energy $\mathcal{F}$ with respect to the magnetization at rest, which
has contributions from the applied, anisotropy, and exchange magnetic fields:
\begin{equation}
\mathbf{H}_{\mathrm{eff}}=-\frac{\delta\mathcal{F}\left[  \mathbf{m}\right]
}{M_{s}\delta\mathbf{m}\left(  \mathbf{r}\right)  }=(H_{\mathrm{ext}}%
+Km_{x})\mathbf{x}-K_{\perp}m_{z}\mathbf{z}+A_{\mathrm{ex}}\nabla
^{2}\mathbf{m}\,,
\end{equation}
where $\mathbf{m}\equiv(m_{x},m_{y},m_{z})=(\cos\theta,\sin\theta\cos\phi
,\sin\theta\sin\phi)$, the anisotropy constants $K>0$, $K_{\perp}>0,$ and the
exchange stiffness $A_{\mathrm{ex}}\ $have been introduced. In the absence of
pinning the Walker ansatz\cite{Schryer}
\begin{equation}
\ln\tan\frac{\theta(x,t)}{2}=-\frac{x-r_{w}(t)}{\lambda_{w}}\,\,\,\mathrm{and}%
\,\,\,\phi\left(  x,t\right)  =\phi(t)\,, \label{Walker}%
\end{equation}
provides a solution in terms of a domain wall with time-dependent position
$r_{w}$ and (squared) width $\lambda_{w}^{2}=A_{\mathrm{ex}}/(K+K_{\perp}%
\sin^{2}\phi).$ The polar angle $\phi$ is the tilt of the magnetization
against the easy-plane anisotropy $K_{\perp}$, which vanishes at equilibrium.
$\phi$ is a constant for sufficiently small, steady-state driving forces, and
therefore not a dynamic variable in the regimes considered henceforth.
Substituting Eq.~(\ref{Walker}) into the Landau-Lifshitz-Gilbert
Eq.~(\ref{LLG})
\begin{equation}
\dot{r}_{w}=\frac{\lambda_{w}}{\alpha}(\dot{\varphi}-\gamma H_{\mathrm{ext}%
})\,,\;K_{\perp}\sin2\phi=-\frac{2(\dot{\varphi}-\gamma H_{\mathrm{ext}}%
)}{\alpha\gamma}\,. \label{rdot}%
\end{equation}
This solution is valid up to a critical (Walker) threshold field at which
$|\sin2\phi_{W}|=1$. To linear order in the driving field, we can approximate
the domain wall width $\lambda_{w}$ by its equilibrium value, $\lambda
_{w}=\sqrt{A_{\mathrm{ex}}/K}$.

The mechanical rotation of the wire is governed by the damped oscillator
equation:
\begin{equation}
I\ddot{\varphi}+\beta^{\mathrm{mech}}\dot{\varphi}=\tau^{\mathrm{mech}},
\end{equation}
where $\beta^{\mathrm{mech}}$ is the mechanical damping parameter and
$\tau^{\mathrm{mech}}$ the total mechanical torque acting along the $x$-axis.
The total angular momentum $\mathfrak{L}_{\mathrm{axis}}$ of the mechanical
and magnetic subsystems in a freely rotating wire of cross section $A$
\begin{equation}
\mathfrak{L}_{\mathrm{axis}}=-\frac{AM_{s}}{\gamma}\left(  l-2r_{w}\right)
+I\dot{\varphi}\,,
\end{equation}
is dissipated into the environment at a rate $\mathfrak{\dot{L}}%
_{\mathrm{axis}}=-\beta^{\mathrm{mech}}\dot{\varphi}$. This leads to an
expression for the Einstein-de Haas torque induced by a moving domain wall:
\begin{equation}
\tau_{\mathrm{EdH}}^{\mathrm{mech}}=-\frac{2AM_{s}}{\gamma}\dot{r}_{w}.
\end{equation}
We assume in the following that the system is overdamped, \textit{i.e}., we
limit our attention to frequencies smaller than $\beta/I$, such that the
acceleration $\ddot{\varphi}$ and moment of inertia $I$ drop out of the
problem. The rotation velocity is then directly proportional to the total
torque $\tau^{\mathrm{mech}}=\tau_{\mathrm{ext}}^{\mathrm{mech}}%
+\tau_{\mathrm{EdH}}^{\mathrm{mech}}$:
\begin{equation}
\beta^{\mathrm{mech}}\dot{\varphi}=\tau_{\mathrm{ext}}^{\mathrm{mech}}%
-\frac{2AM_{s}}{\gamma}\dot{r}_{w}\,. \label{mech}%
\end{equation}
The mechanical energy $E(\varphi)$ governs the external torque, $\tau
_{\mathrm{ext}}^{\mathrm{mech}}\equiv-\partial_{\varphi}E(\varphi)$.

The above results will now be shown to be consistent with Onsager's
reciprocity principle and the second law of thermodynamics. Disregarding
thermal effects, it is natural to switch to the free energy $\mathcal{F}$
instead of the entropy $\mathcal{S}$:
\begin{equation}
\mathcal{F}(r_{w},\varphi)=\mathcal{F}_{w}+\mathcal{F}_{\varphi}%
=(2r_{w}-l)AM_{s}H_{\mathrm{ext}}+E(\varphi)\,, \label{freeE}%
\end{equation}
where $l$ is the total length of the wire and the domain wall position $r_{w}$
is measured with respect to the left end of the wire. We omit the internal
energy of the domain wall, which below the Walker threshold may be treated as
a rigid particle-like mass-less object specified by its position. The
conjugate forces associated with $r_{w}$ and $\varphi$ are immediately found
as
\begin{equation}
X_{w}\equiv-\frac{\partial}{\partial r_{w}}\mathcal{F}=-2AM_{s}H_{\mathrm{ext}%
}\,;\;X_{\varphi}\equiv-\frac{\partial}{\partial\varphi}\mathcal{F}%
=\tau_{\mathrm{ext}}^{\mathrm{mech}}\,.
\end{equation}
After simple algebra using Eqs.~(\ref{rdot}) and (\ref{mech}), the energy
dissipation is found to be positive definite:
\begin{equation}
T\dot{\mathcal{S}}\equiv-\dot{\mathcal{F}}=-2AM_{s}H_{\mathrm{ext}}\dot{r}%
_{w}+\tau_{\mathrm{ext}}^{\mathrm{mech}}\dot{\varphi}=\frac{2\alpha AM_{s}%
}{\gamma\lambda_{w}}\dot{r}_{w}^{2}+\beta^{\mathrm{mech}}\dot{\varphi}^{2}%
\geq0\,,
\end{equation}

Rewriting the equations of motion (\ref{rdot}) and (\ref{mech}), the cross
terms are seen to obey Onsager's symmetry:
\begin{equation}
\left(  1+\frac{2AM_{s}}{\gamma}\frac{\lambda_{w}}{\alpha\beta^{\mathrm{mech}%
}}\right)  \left(
\begin{array}
[c]{c}%
\dot{\varphi}\\
\dot{r}_{w}%
\end{array}
\right)  =\left(
\begin{array}
[c]{cc}%
\frac{1}{\beta^{\mathrm{mech}}} & -\frac{\lambda_{w}}{\alpha\beta
^{\mathrm{mech}}}\\
\frac{\lambda_{w}}{\alpha\beta^{\mathrm{mech}}} & \frac{\lambda_{w}\gamma
}{2\alpha AM_{s}}%
\end{array}
\right)  \left(
\begin{array}
[c]{c}%
X_{\varphi}\\
X_{w}%
\end{array}
\right)  \ \label{Mech1}%
\end{equation}
The antisymmetry of the off-diagonal terms stems from Onsager's reciprocity,
which relates here the response of the tail-to-tail domain wall to that of its
time-reversed partner, which is a head-to-head domain wall. Note that the
inverse of Eq. (\ref{Mech1}) is simpler
\begin{equation}
\left(
\begin{array}
[c]{c}%
X_{\varphi}\\
X_{w}%
\end{array}
\right)  =\left(
\begin{array}
[c]{cc}%
\beta^{\mathrm{mech}} & \frac{2AM}{\gamma}\\
-\frac{2AM}{\gamma} & \frac{2AM}{\gamma}\frac{\alpha}{\lambda}%
\end{array}
\right)  \left(
\begin{array}
[c]{c}%
\dot{\varphi}\\
\dot{r}_{w}%
\end{array}
\right)  ,\
\end{equation}
but Eq. (\ref{Mech1}) should be closer to experimental set-ups in practice. We
may rewrite it as%
\begin{equation}
\left(
\begin{array}
[c]{c}%
\dot{\varphi}\\
\dot{r}_{w}%
\end{array}
\right)  =\left(
\begin{array}
[c]{cc}%
\frac{1}{\tilde{\beta}^{\mathrm{mech}}} & -\frac{\lambda_{w}}{\tilde{\alpha
}\beta^{\mathrm{mech}}}\\
\frac{\lambda_{w}}{\tilde{\alpha}\beta^{\mathrm{mech}}} & \frac{\lambda
_{w}\gamma}{2\tilde{\alpha}AM_{s}}%
\end{array}
\right)  \left(
\begin{array}
[c]{c}%
X_{\varphi}\\
X_{w}%
\end{array}
\right)  \,,
\end{equation}
where%
\begin{align}
\tilde{\beta}^{\mathrm{mech}}  &  =\beta^{\mathrm{mech}}+\frac{2\lambda
_{w}AM_{s}}{\alpha\gamma}\,,\nonumber\\
\tilde{\alpha}  &  =\alpha+\frac{2\lambda_{w}AM_{s}}{\gamma\beta
^{\mathrm{mech}}}\,. \label{alphatilde}%
\end{align}
The magnetomechanical coupling creates an apparently increased damping of the
magnetization dynamics and/or the mechanical motion that is proportional to
the number of spins in the domain wall. $\beta^{\mathrm{mech}}$ is the
mechanical friction: when it becomes large the mechanical motion is quenched
and the excess Gilbert damping is suppressed $\tilde{\alpha}\rightarrow\alpha
$. In turn, the direct coupling of the mechanical torque to the rotation,
$L_{\varphi\varphi}=1/\tilde{\beta}^{\mathrm{mech}}$, vanishes with vanishing
Gilbert damping $\alpha$, \textit{i.e}., $\tau_{\mathrm{ext}}^{\mathrm{mech}}$
is fully transferred into the magnetic system. For vanishing mechanical
damping $\beta^{\mathrm{mech}}$, the domain wall remains immobile under a
magnetic field, but the wire rotates with an $\alpha$-independent angular
velocity, which exactly compensates the external field in the rotating frame
(i.e., $\dot{\varphi}=\gamma H_{\mathrm{ext}}$). These results are valid only
in the steady-state, overdamped mechanical regime considered here.

\section{Magnetomechanothermoelectric systems}

We now define the conjugate thermodynamical variables $a_{i}$ that allow us to
take advantage of Onsager's relations as energy transfer $U=(U_{2}-U_{1})/2$,
charge transfer $q=(q_{2}-q_{1})/2$, domain wall position $r_{w}$, and
lattice-rotation angle $\varphi.$ The corresponding fluxes are given by their
time derivatives $J_{Q}=\dot{U},$ $J_{c}=\dot{q},$ $J_{w}=\dot{r}_{w},$
$J_{\varphi}=\dot{\varphi}$. The thermodynamic forces (\ref{forces}) depend in
principle on the values of all thermodynamic variables. It is possible to work
out a general scheme that includes all possible cross correlations, but it
would not be very transparent. Instead, we follow a more pragmatic approach
that is based on the low-temperature free energy for the magnetomechanical
degrees of freedom, which are coupled to thermoelectric transport between the
reservoirs by the spin torques. The linear response matrix then reads
$\mathbf{J}=\hat{L}\mathbf{X}$, where%
\begin{align}
\mathbf{J}  &  =\left(
\begin{array}
[c]{cccc}%
J_{c}, & J_{Q}, & \dot{\varphi}, & \dot{r}_{w}%
\end{array}
\right)  ^{T}\\
\mathbf{X}  &  =\left(
\begin{array}
[c]{cccc}%
-\Delta V, & -\frac{\Delta T}{T}, & \tau_{\mathrm{ext}}^{\mathrm{mech}}, &
-2AM_{s}H_{\mathrm{ext}}%
\end{array}
\right)  ^{T}%
\end{align}
and%
\begin{equation}
\hat{L}=\left(
\begin{array}
[c]{cccc}%
L_{cc} & L_{cQ} & L_{c\varphi} & L_{cw}\\
L_{Qc} & L_{QQ} & L_{Q\varphi} & L_{Qw}\\
L_{\varphi c} & L_{\varphi Q} & L_{\varphi\varphi} & L_{\varphi w}\\
L_{wc} & L_{wQ} & L_{w\varphi} & L_{ww}%
\end{array}
\right)  . \label{Onsager}%
\end{equation}
When thermoelectric and magnetomechanical systems are uncoupled, the matrix
elements derived in Sec. II may be filled in unmodified. According to the
Onsager symmetry, $L_{xw}(\mathbf{m})=L_{wx}(-\mathbf{m})=L_{wx}(\mathbf{m})$
and $L_{x\varphi}(\mathbf{m})=L_{\varphi x}(-\mathbf{m})=-L_{\varphi
x}(\mathbf{m})$, for $x=(c,Q)$, assuming that our system obeys a structural
mirror symmetry with respect to a plane normal to the wire in Fig.~1. It is
useful to introduce also the inverse matrix $\mathbf{X}$ $=\hat{L}%
^{-1}\mathbf{J}$, recalling that $\hat{L}^{-1}$ and $\hat{L}$ have the same
Onsager symmetry.

We can draw a number of conclusions from the Onsager relations already.
$L_{wc}$ and $L_{cw}$ represent the Onsager equivalent pair of current-induced
transfer torque and charge pumping by the magnetization dynamics,
respectively.\cite{Saslow,Duine,Mecklenburg} We know that a temperature
gradient can induce a spin-transfer torque,\cite{Hatami} which is here
represented by $L_{wQ}.$ According to Onsager symmetry an opposite and
equivalent effect exists, \textit{i.e}., a heat current induced by
magnetization dynamics, which might be applied for cooling or heating
purposes. As explained above, the mechanical motion induced by the magnetic
field as quantified by $L_{\varphi w}$ (Einstein-de Haas effect) is identical
with the Barnett response function $-L_{w\varphi},$ which describes the
magnetization dynamics induced by rotation (Barnett effect). Since $L_{\varphi
c}=-L_{c\varphi}$ the magnetic wire can be employed as an
electromotor\cite{Kovalev03} and electric generator. A temperature gradient
induces a rotation of the wire via $L_{\varphi Q}$, which leads to the
prediction of a heat engine that can carry out mechanical work under a
temperature difference. The opposite effect, in which mechanical motion of the
wire is transformed into a temperature gradient is governed by $L_{Q\varphi
}=-L_{\varphi Q}$.

The remaining task is to work out the elements of the $4\times4$ response
matrix. In the adiabatic regime, the magnetic texture varies slowly with
respect to the magnetic coherence length $1/\lambda_{c}=|1/\lambda
_{F}^{\downarrow}-1/\lambda_{F}^{\uparrow}|,$ where $\lambda_{F}%
^{\uparrow\left(  \downarrow\right)  }$ are the spin-dependent Fermi
wavelengths and $\uparrow/\downarrow$ denote the majority/minority spin
carriers, respectively. The spin torque on, or angular momentum transfer to,
the magnetization induced by an applied voltage (superscript $\left(
0\right)  $ indicates a static magnetization texture and the absence of
thermoelectric effects) reads
\cite{Zhang,Thiaville,Mougin,Tatara,Tserkovnyak08}
\begin{equation}
\left(  \mbox{\boldmath$\tau$}_{c}^{\mathrm{mag}}\right)  ^{\left(  0\right)
}=-\frac{\hbar}{e}\frac{\gamma}{2AM_{s}}PG\Delta V\left(  1-\beta
_{c}^{\mathrm{mag}}\mathbf{m}\times\right)  \frac{\partial}{\partial
x}\mathbf{m}\,. \label{tauC0}%
\end{equation}
in terms of the spin polarization $P=\left(  G_{\uparrow}-G_{\downarrow
}\right)  /G$ of the electric conductance $G=G_{\uparrow}+G_{\downarrow}$ of
the single-domain ferromagnet. The prefactor (torque in the plane of the
domain wall magnetization) can be easily identified as the angular momentum
rate of change of the spin-polarized carriers that corresponds to an adiabatic
spin reversal. The out-of-plane torque components is caused by the mistracking
of the spin in the magnetization texture that is parameterized by $\beta
_{c}^{\mathrm{mag}}$. The torque by the thermoelectric spin current induced by
a temperature bias (superscript $\left(  0\right)  $ again denoting a static
texture and the condition $S^{2}/\mathcal{L}\ll1)$ reads analogously:
\begin{equation}
\left(  \mbox{\boldmath$\tau$}_{Q}^{\mathrm{mag}}\right)  ^{\left(  0\right)
}=-\frac{\hbar}{e}\frac{\gamma}{2AM_{s}}P^{\prime}SG\Delta T\left(
1-\beta_{Q}^{\mathrm{mag}}\mathbf{m}\times\right)  \frac{\partial}{\partial
x}\mathbf{m}\,. \label{tauQ0}%
\end{equation}
where $P^{\prime}=\partial_{E}\left(  PG\right)  /\partial_{E}G$ is the
polarization of the energy derivative of the conductance.\cite{Note} The
parameter $\beta_{Q}^{\mathrm{mag}}$ parameterizing the out-of-plane torque
component differs from $\beta_{c}^{\mathrm{mag}}$ since the non-equilibrium
energy distribution defining the spin current through the texture has a node
at the Fermi energy rather than a maximum. The Seebeck coefficient for the
homogeneous ferromagnet $S=\left(  S_{\uparrow}G_{\uparrow}+S_{\downarrow
}G_{\downarrow}\right)  /G.$

In the coupled system, torques are induced by the magnetization and mechanical
motion as well, which can be fully included into the equations of motion by
adapting the charge and heat currents, rather than voltage and temperature as
system variables (forces). The transformation can be carried out very
generally, but leads to different parameters for the out-of-plane
torques.\cite{Kovalev09} By inverting the thermoelectric matrix in the
Sommerfeld approximation and assuming $S^{2}/\mathcal{L}\ll1,$ the $\beta$
parameters remain unmodified for the current biased torques, however. The
torque induced by a heat current $J_{Q}$ then reads
\begin{equation}
\mbox{\boldmath$\tau$}_{Q}^{\mathrm{mag}}=\frac{\hbar}{e}\frac{\gamma}%
{2AM_{s}}\frac{P^{\prime}S}{\mathcal{L}T}J_{Q}\left(  1-\beta_{Q}%
^{\mathrm{mag}}\mathbf{m}\times\right)  \frac{\partial}{\partial x}%
\mathbf{m}\,, \label{tauC}%
\end{equation}
whereas the charge current torque becomes%
\begin{equation}
\mbox{\boldmath$\tau$}_{c}^{\mathrm{mag}}=\frac{\hbar}{e}\frac{\gamma}%
{2AM_{s}}P\left(  J_{c}-\frac{S}{\mathcal{L}T}J_{Q}\right)  \left(
1-\beta_{c}^{\mathrm{mag}}\mathbf{m}\times\right)  \frac{\partial}{\partial
x}\mathbf{m}\,. \label{tauQ}%
\end{equation}
Note that the conventional thermoelectric charge current has been subtracted
here from the total charge current. Adding the spin torques (\ref{tauC}) and
(\ref{tauQ}) to the right-hand side of the LLG Eq.~(\ref{LLG}), we can employ
the Walker ansatz again to solve for the current driven domain-wall velocity:
\begin{equation}
\dot{r}_{w}|_{J_{c},J_{Q}}=-\frac{\hbar G}{e}\frac{\gamma}{2AM_{s}}\frac
{1}{\alpha}\left(  P\beta_{c}^{\mathrm{mag}}\left(  J_{c}-\frac{S}%
{\mathcal{L}T}J_{Q}\right)  +\frac{P^{\prime}S}{\mathcal{L}T}\beta
_{Q}^{\mathrm{mag}}J_{Q}\right)  . \label{jdw}%
\end{equation}
A negative charge, thus positive particle, current and $P\beta_{c}%
^{\mathrm{mag}}>0$ pushes the domain wall to the right. For an electron-like
thermopower ($S<0$) and $P^{\prime}\beta_{Q}^{\mathrm{mag}}>0$, a positive
heat current has the same effect.

In order to relate $\dot{\varphi}$ to the mechanical torque and identify the
unknown response coefficients in Eq.~(\ref{Onsager}), we need to generalize
the conservation of angular momentum, Eq.~(\ref{mech}), to account for the
spin currents injected into and drained from the leads by
\begin{equation}
\beta^{\mathrm{mech}}\dot{\varphi}=\tau_{\mathrm{ext}}^{\mathrm{mech}}%
-\frac{2AM_{s}}{\gamma}\dot{r}_{w}+\tau_{\mathrm{inj}}^{\mathrm{mech}}\,,
\label{meqm}%
\end{equation}
with, for the present choice of magnetization texture,%
\begin{equation}
\tau_{\mathrm{inj}}^{\mathrm{mech}}=-\frac{\hbar}{e}(\gamma_{c}P\left(
J_{c}-\frac{S}{\mathcal{L}T}J_{Q}\right)  +\gamma_{Q}\frac{P^{\prime}%
S}{\mathcal{L}T}J_{Q}). \label{tauin}%
\end{equation}
The effect is maximized ($\gamma_{c}=\gamma_{Q}=1$) when the angular momentum
is drained completely from the wire into the reservoirs. In the opposite
limit, the spin currents are dissipated completely in the wire (rather than in
the reservoir), as in sufficiently long normal metal terminals to the
ferromagnet (N$|$F$|$N) that are part of the mounted wire, such that
$\tau_{\mathrm{inj}}^{\mathrm{mech}}=0$. The domain wall equation of motion,
including the Barnett torques induced by rotation, reads
\begin{equation}
\dot{r}_{w}=-\frac{\lambda_{w}\gamma}{\alpha}\left(  H_{\mathrm{ext}}%
-\frac{\dot{\varphi}}{\gamma}\right)  +\dot{r}_{w}|_{J_{c},J_{Q}}\,.
\label{mm}%
\end{equation}
From Eqs.~(\ref{jdw},\ref{meqm},\ref{tauin},\ref{mm}) we can specify all
coefficients of the inverse response matrix%
\begin{equation}
\hat{L}^{-1}=\left(
\begin{array}
[c]{cccc}%
\frac{1}{\kappa}\left(  TS^{2}+\frac{\kappa}{G}\right)  & -\frac{S}{\kappa} &
-\frac{\hbar}{e}P\gamma_{c} & \frac{\hbar}{\lambda_{w}e}P\beta_{c}%
^{\mathrm{mag}}\\
-\frac{S}{\kappa} & \frac{1}{T\kappa} & -\frac{\hbar}{e}\left(  P^{\prime
}\gamma_{Q}-P\gamma_{c}\right)  \frac{S}{\mathcal{L}T} & \frac{\hbar}{e}%
\frac{S}{\lambda_{w}\mathcal{L}T}\left(  P^{\prime}\beta_{Q}^{\mathrm{mag}%
}-P\beta_{c}^{\mathrm{mag}}\right) \\
\frac{\hbar}{e}P\gamma_{c} & \frac{\hbar}{e}\frac{S}{\mathcal{L}T}\left(
P^{\prime}\gamma_{Q}-P\gamma_{c}\right)  & \beta^{\mathrm{mech}} &
\frac{2AM_{s}}{\gamma}\\
\frac{\hbar}{\lambda_{w}e}P\beta_{c}^{\mathrm{mag}} & \frac{\hbar}{e}\frac
{S}{\lambda_{w}\mathcal{L}T}\left(  P^{\prime}\beta_{Q}^{\mathrm{mag}}%
-P\beta_{c}^{\mathrm{mag}}\right)  & -\frac{2AM_{s}}{\gamma} & \frac
{2AM_{s}\alpha}{\gamma\lambda_{w}}%
\end{array}
\right)  \label{Onsagerinv}%
\end{equation}

This representation appears to be less convenient for comparison with
practical experiment. In a purely electric circuit it is possible to freely
change from a current-biased to a voltage bias set-up. This appears less
convenient for the other sets of conjugate variables. It is therefore
necessary to adopt the results to the experimental problem at hand. For a
set-up in which the driving forces are the $X_{i}$ considered here, it is
appropriate to invert the above matrix in order to obtain experimentally more
relevant response functions. We have seen in the previous section that for the
purely magnetomechanical system the inversion is equivalent to a
renormalization of the damping constants. The inversion of the $4\times4$
matrix leads to lengthy expressions that cannot be interpreted that easily.
The simplest approach is second order perturbation theory to estimate the
importance of the self-consistent couplings. The diagonal elements of the
response matrix then read%
\begin{equation}
L_{ii}\approx\frac{1}{\left(  L^{-1}\right)  _{ii}}\left(  1+\sum_{j\neq
i}\frac{\left(  L^{-1}\right)  _{ij}\left(  L^{-1}\right)  _{ji}}{\left(
L^{-1}\right)  _{ii}\left(  L^{-1}\right)  _{jj}}\right)  ,
\end{equation}
while the non-diagonal elements become%

\begin{equation}
L_{ij}\approx\frac{\left(  L^{-1}\right)  _{ji}}{\left(  L^{-1}\right)
_{ii}\left(  L^{-1}\right)  _{jj}}%
\end{equation}

In Eq. (\ref{Onsagerinv}) the $2\times2$ thermoelectric matrix and the
mechanical diagonal elements scale with the system length and inversely with
the wire cross section $l/A$, whereas all others are independent of the system
size. The non-diagonal block matrices may therefore be treated by perturbation
theory in the long and/or narrow wire limit. By defining the block-diagonal
matrix
\begin{equation}
\hat{L}_{0}^{-1}=\left(
\begin{array}
[c]{cccc}%
\frac{1}{\kappa}\left(  TS^{2}+\frac{\kappa}{G}\right)  & -\frac{S}{\kappa} &
0 & 0\\
-\frac{S}{\kappa} & \frac{1}{T\kappa} & 0 & 0\\
0 & 0 & \beta^{\mathrm{mech}} & \frac{2AM_{s}}{\gamma}\\
0 & 0 & -\frac{2AM_{s}}{\gamma} & \frac{2AM_{s}}{\gamma}\frac{\alpha}{\lambda}%
\end{array}
\right)  .
\end{equation}
and treating $\delta\hat{L}^{-1}=$ $\hat{L}^{-1}-\hat{L}_{0}^{-1}\ $ as a
perturbation, we find to lowest order in $\delta\hat{L}^{-1}$%
\begin{equation}
\hat{L}\approx\hat{L}_{0}-\hat{L}_{0}\delta\hat{L}^{-1}\hat{L}_{0}.
\end{equation}
Using the Sommerfeld approximation and letting $S^{2}/\mathcal{L}\rightarrow
0$, we obtain the elements of the lower non-diagonal block as:%
\begin{equation}
L_{wc}=-\frac{\hbar}{e}\frac{\gamma}{2AM_{s}}\frac{G}{\tilde{\alpha}}P\left(
\beta_{c}^{\mathrm{mag}}+\frac{2AM_{s}}{\gamma}\frac{\lambda_{w}}%
{\beta^{\mathrm{mech}}}\gamma_{c}\right)  ,
\end{equation}%
\begin{equation}
L_{wQ}=-\frac{\hbar}{e}\frac{\gamma}{2AM_{s}}\frac{GST}{\tilde{\alpha}%
}P^{\prime}\left(  \beta_{Q}^{\mathrm{mag}}+\frac{2AM_{s}}{\gamma}%
\frac{\lambda_{w}}{\beta^{\mathrm{mech}}}\gamma_{Q}\right)  ,
\end{equation}%
\begin{equation}
L_{\varphi c}=-\frac{\hbar}{e}\frac{G}{\tilde{\beta}^{\mathrm{mech}}}P\left(
\gamma_{c}-\frac{\beta_{c}^{\mathrm{mag}}}{\alpha}\right)  ,
\end{equation}%
\begin{equation}
L_{\varphi Q}=-\frac{\hbar}{e}\frac{GST}{\tilde{\beta}^{\mathrm{mech}}%
}P^{\prime}\left(  \gamma_{Q}-\frac{\beta_{Q}^{\mathrm{mag}}}{\alpha}\right)
.
\end{equation}
The perturbation expansion holds well for the example treated below. If it
turns out inaccurate, the full matrix should be diagonalized, of course.

In strongly spin-orbit coupled systems the transport polarizations
$P$,$P^{\prime}$ are not well defined.\cite{Kovalev09} However, the dynamics
is governed only by the combinations $P\beta_{c}^{\mathrm{mag}},P\gamma
_{c},P^{\prime}\beta_{Q}^{\mathrm{mag}},P^{\prime}\gamma_{Q}$, which can still
be determined.\cite{Hals3} An important handle to experimental access to the
different parameters is a comparison of heat-current driven domain wall motion
for both closed and open electric circuits. In the limit $1/\beta
^{\mathrm{mech}}=H_{\mathrm{ext}}=0$, the former case leads to
\begin{equation}
\dot{r}_{w}|_{\Delta V=0}=-\frac{\gamma}{2AM_{s}}\frac{\hbar GST}{e}%
\frac{P^{\prime}\beta_{Q}^{\mathrm{mag}}}{\alpha}\left(  -\frac{\Delta T}%
{T}\right)
\end{equation}
whereas
\begin{equation}
\dot{r}_{w}|_{J_{c}=0}=-\frac{\gamma}{2AM_{s}}\frac{\hbar GST}{e}%
\frac{P^{\prime}\beta_{Q}^{\mathrm{mag}}-P\beta_{c}^{\mathrm{mag}}}{\alpha
}\left(  -\frac{\Delta T}{T}\right)
\end{equation}

An interesting simplified system consists of a completely pinned magnetic
domain wall. In this regime, the magnetic degrees of freedom drop out of the
problem.\cite{Note2} In the adiabatic limit, the spin current then transfers
all angular momentum directly to the lattice. Since the magnetization does not
move, there is no magnetic dissipation. The response functions in that limit
(indicated by a prime) are obtained in the limit $\alpha\rightarrow\infty$ and
are significantly simplified%
\begin{equation}
L_{\varphi c}^{\prime}=-\frac{\hbar}{e}\frac{G}{\beta^{\mathrm{mech}}}%
P\gamma_{c},
\end{equation}%
\begin{equation}
L_{\varphi Q}^{\prime}=-\frac{\hbar}{e}\frac{GST}{\beta^{\mathrm{mech}}%
}P^{\prime}\gamma_{Q}.
\end{equation}
The Onsager equivalent to the current-induced rotation is the rotation-induced
charge and heat pumping by the otherwise fixed magnetization texture in the
domain wall. It can be explained in terms of the magnetization texture that
carries out a rotation rather than a translation, which in the rotating frame
results in an effective (Barnett-like) field $\dot{\varphi}/\gamma$ between
the two reservoirs, which drives the charge and heat currents. Whether the
pinned or the moving magnetization more effectively transfer angular momentum
between currents and lattice depends strongly on the ratio of the dissipative
out-of-plane torques and the Gilbert damping constant. The situation in real
domain walls with weak pinning will be somewhere between the extremes of rigid
translation and full pinning, but its full treatment is beyond the scope of
the present work.

The dynamics of insulating ferromagnets can be obtained by simply crossing out
the first row and column of Eq.~(\ref{Onsagerinv}) related to the charge
degree of freedom. In the remaining $3\times3$ matrix, the spin torques are
exerted by the pure heat currents carried by spin waves, unlike in metallic
systems, in which the spin torque is dominated by the electric current. The
detailed response function for insulating ferromagnets will be discussed
separately, however.

\section{Scattering theory}

The magnetic damping and the charge-current magnetization coupling have been
determined microscopically by scattering theory.\cite{Brataas08,Hals,Starikov}
Here we briefly review the relevant published results and add new ones related
to heat transport.

The Onsager response functions derived above contain a number of parameters,
basically the spin-dependent conductances at the Fermi energy $G_{\sigma
}=G_{\sigma}\left(  E_{F}\right)  ,$ the Gilbert damping $\alpha$ and the
dissipative out-of-plane spin transfer torque associated to charge current
$\beta_{c}^{\mathrm{mag}}$ and heat current $\beta_{Q}^{\mathrm{mag}}.$ They
can all be written in terms of the scattering matrix $\hat{S}$ of the wire at
a given energy.\cite{Brataas08,Hals} Using the conventional notation in terms
of transmission ($\hat{t},\hat{t}^{\prime}$) and reflection ($\hat{r},\hat
{r}^{\prime}$) matrices\cite{Nazarov} the scattering matrix in the space of
the transport channels to and from the wire at an energy $E$ and spin indices
$\sigma,\sigma^{\prime}$ reads:%
\begin{equation}
\hat{S}_{\sigma\sigma^{\prime}}\left(  E\right)  =\left(
\begin{array}
[c]{cc}%
\hat{r}_{\sigma\sigma^{\prime}}\left(  E\right)  & \hat{t}_{\sigma
\sigma^{\prime}}^{\prime}\left(  E\right) \\
\hat{t}_{\sigma\sigma^{\prime}}\left(  E\right)  & \hat{r}_{\sigma
\sigma^{\prime}}^{\prime}\left(  E\right)
\end{array}
\right)  .
\end{equation}
The spin-dependent conductance of the (single-domain) ferromagnet can be
expressed by the Landauer-B\"{u}ttiker formula:
\begin{equation}
G_{\sigma}=\frac{e^{2}}{h}\mathrm{Tr}\sum_{\sigma^{\prime}}\;\hat{t}%
_{\sigma\sigma^{\prime}}^{\dag}\hat{t}_{\sigma\sigma^{\prime}},
\end{equation}
where the trace indicates the sum over (orbital) transport channels at the
Fermi energy. The Mott formula for the conventional\ thermopower
$S=-e\mathcal{L}T\partial_{E}\ln G$ can be computed from the energy-dependent
conductance $G=G_{\uparrow}+G_{\downarrow}$.

When $\hat{S}$ is the scattering matrix of the ferromagnet including one
domain wall at $r_{w}$, the parametric pumping of a charge
current\cite{Buttiker} by the moving domain wall\cite{Hals} into the right
leads reads%
\begin{equation}
J_{c,w}=-\frac{e}{4\pi}\dot{r}_{w}\operatorname{Im}\mathrm{Tr}_{s}%
\frac{\partial\hat{S}}{\partial r_{w}}\hat{S}^{\dag}\hat{\tau}_{z},
\label{Jcw}%
\end{equation}
where
\begin{equation}
\hat{\tau}_{z}=\left(
\begin{array}
[c]{cc}%
\hat{1} & 0\\
0 & -\hat{1}%
\end{array}
\right)
\end{equation}
in the same space as the scattering matrix, \textit{i.e}. propagating states
in the left lead and the right lead, respectively, and $\mathrm{Tr}_{s}$ is
the sum over these states (including spin). The expression for the energy
pumped out of the system with a parametric time dependence of the scattering
matrix\cite{Avron,Moskalets}
\begin{equation}
J_{E}=\frac{\hbar}{4\pi}\mathrm{Tr}_{s}\frac{\partial\hat{S}}{\partial t}%
\frac{\partial\hat{S}^{\dag}}{\partial t}=\frac{\hbar}{4\pi}\left(  \dot
{r}_{w}\right)  ^{2}\mathrm{Tr}_{s}\frac{\partial\hat{S}}{\partial r_{w}}%
\frac{\partial\hat{S}^{\dag}}{\partial r_{w}} \label{JE}%
\end{equation}
has been employed by Brataas \textit{et al}.\cite{Brataas08} to derive
microscopic expressions for the Gilbert damping and by Hals \textit{et
al}.\cite{Hals} for the charge current-induced domain wall motion. When
evaluating the scattering matrix for zero bias and assuming that the domain
wall is driven by a magnetic field $J_{E}=\left(  \dot{r}_{w}\right)
^{2}/L_{ww}$:
\begin{equation}
L_{ww}=\left(  \frac{\hbar}{4\pi}\mathrm{Tr}_{s}\frac{\partial\hat{S}%
}{\partial r_{w}}\frac{\partial\hat{S}^{\dag}}{\partial r_{w}}\right)  ^{-1}
\label{LwwS}%
\end{equation}
For $S^{2}/\mathcal{L}\ll1$ and absence of rotation the response function
reads to lowest order in the conductance
\begin{equation}
L_{ww}\approx\frac{\gamma\lambda_{w}}{2AM_{s}\alpha}+G\left(  \frac{\hbar}%
{e}\frac{\gamma P\beta_{c}^{\mathrm{mag}}}{2AM_{s}\alpha}\right)  ^{2},
\label{LwwExp}%
\end{equation}%
\begin{equation}
L_{wc}\approx-\frac{\hbar}{e}\frac{\gamma}{2AM_{s}}\frac{G}{\alpha}P\beta
_{c}^{\mathrm{mag}},
\end{equation}
For long wire lengths $l$ we recover the result by Hals \textit{et
al.}\cite{Hals}\textit{ }for the Gilbert damping\textit{:}
\begin{equation}
\alpha=\frac{\gamma\hbar\lambda_{w}}{8\pi AM_{s}}\lim_{l\rightarrow\infty
}\mathrm{Tr}_{s}\frac{\partial\hat{S}}{\partial r_{w}}\frac{\partial\hat
{S}^{\dag}}{\partial r_{w}}.
\end{equation}
and the dissipative torque correction%
\begin{equation}
P\beta_{c}^{\mathrm{mag}}=\frac{e^{2}\lambda_{w}}{2h}\lim_{l\rightarrow\infty
}\frac{1}{G}\operatorname{Im}\mathrm{Tr}_{s}\frac{\partial\hat{S}}{\partial
r_{w}}\hat{S}^{\dag}\hat{\tau}_{z}.
\end{equation}

The heat current pumped by the magnetization dynamics depends linearly on the
frequency and amplitude of the pumping parameter and should not be confused
with the energy current $J_{E}$,$\ $Eq. (\ref{JE}), which is to leading order
quadratic in these quantities. This thermoelectric contribution to the pumping
current can be obtained by a Sommerfeld expansion of the energy dependent
parametric pumping current as derived by Moskalet and
B\"{u}ttiker.\cite{Moskalets} The heat current driven by a moving domain wall
then reads
\begin{equation}
J_{Q}=-e\mathcal{L}T^{2}\frac{\partial}{\partial E}J_{c,w}\left(  E\right)  .
\end{equation}
where $J_{c,w}$ is a function of energy. Observing that the domain wall
velocity in Eq. (\ref{Jcw}) is a parameter that can be pulled in front of the
energy-derivative we arrive at
\begin{equation}
J_{Q}=-\frac{e}{4\pi}\dot{r}_{w}\frac{\partial}{\partial E}\operatorname{Im}%
\mathrm{Tr}_{s}\frac{\partial\hat{S}}{\partial r_{w}}\hat{S}^{\dag}\hat{\tau
}_{z}.
\end{equation}
This leads to
\begin{equation}
\lim_{l\rightarrow\infty}L_{wQ}=\mathcal{L}T^{2}\frac{e^{2}}{2\hbar}%
\frac{\partial_{E}\operatorname{Im}\mathrm{Tr}_{s}\frac{\partial\hat{S}%
}{\partial r_{w}}\hat{S}^{\dag}\hat{\tau}_{z}\ }{\mathrm{Tr}_{s}\frac
{\partial\hat{S}}{\partial r_{w}}\frac{\partial\hat{S}^{\dag}}{\partial r_{w}%
}} \label{LwQ0fp}%
\end{equation}
In the limit of long wires the leading term of the heat-domain wall coupling
\begin{equation}
\lim_{l\rightarrow\infty}L_{wQ}=-\frac{\hbar}{e}\frac{GST}{\alpha}\frac
{\gamma}{2AM_{s}}P^{\prime}\beta_{Q}^{\mathrm{mag}}.
\end{equation}
and therefore:%
\begin{equation}
P^{\prime}\beta_{Q}^{\mathrm{mag}}=\frac{e^{2}\lambda_{w}}{2h}\lim
_{l\rightarrow\infty}\frac{\partial_{E}\operatorname{Im}\mathrm{Tr}_{s}%
\frac{\partial\hat{S}}{\partial r_{w}}\hat{S}^{\dag}\hat{\tau}_{z}}%
{\partial_{E}G},
\end{equation}
which has been recently evaluated by Hals \textit{et al}. for
GaMnAs.\cite{Hals3} We can also derive a relation between the parameters
governing the charge current and heat current-induced domain wall motion%
\begin{equation}
SP^{\prime}\beta_{Q}^{\mathrm{mag}}=S_{\beta}P\beta_{c}^{\mathrm{mag}},
\end{equation}
where
\begin{equation}
S_{\beta}=-e\mathcal{L}T\frac{\partial}{\partial E}\ln\left(  GP\beta
_{c}^{\mathrm{mag}}\right)  .
\end{equation}

\section{Numerical estimates}

To mount magnetic wires such that they can rotate freely seems challenging,
but should be possible.\cite{Kovalev07} Elias \textit{et al}. have grown
single-crystalline FeCo\ wires inside multi-wall carbon nanotubes.\cite{Elias}
The outer walls of multi-wall carbon nanotubes form almost ideal bearings for
the rotation of the inner tubes.\cite{Fennimore,Servantie} A possible recipe
for creating a system that can be described by the present model is therefore
a suspended bridge of a multi-wall coated FeCo nanowire. In order to insure
that all currents flow through the ferromagnet, it might be useful to burn off
the carbon in the free standing part. Such a system could sustain GHz rotation
frequencies when driven by the spin-flip transfer torque that dissipates an
injected spin current.\cite{Kovalev07} The scaling with different material
constant is obvious in Eq. (\ref{Onsagerinv}). We chose parameters that are
close to permalloy, \textit{viz}. $\rho=10^{-5}%
%TCIMACRO{\unit{\U{3a9}}}%
%BeginExpansion
\operatorname{\Omega }%
%EndExpansion%
%TCIMACRO{\unit{cm}}%
%BeginExpansion
\operatorname{cm}%
%EndExpansion
,\;\lambda_{w}=100%
%TCIMACRO{\unit{nm}}%
%BeginExpansion
\operatorname{nm}%
%EndExpansion
;\;S=-40%
%TCIMACRO{\unit{\U{3bc}V}}%
%BeginExpansion
\operatorname{\mu V}%
%EndExpansion%
%TCIMACRO{\unit{K}}%
%BeginExpansion
\operatorname{K}%
%EndExpansion
^{-1}.$ Servantie and Gaspard\cite{Servantie} report the dynamic friction
$\beta^{\mathrm{mech}}/l=0.044%
%TCIMACRO{\unit{u}}%
%BeginExpansion
\operatorname{u}%
%EndExpansion%
%TCIMACRO{\unit{nm}}%
%BeginExpansion
\operatorname{nm}%
%EndExpansion
/%
%TCIMACRO{\unit{ps}}%
%BeginExpansion
\operatorname{ps}%
%EndExpansion
$ for a $\left(  4,4\right)  $ nanotube rotating in a $\left(  9,9\right)  $
nanotube bearing. We choose a wire area cross section $A=100%
%TCIMACRO{\unit{nm}}%
%BeginExpansion
\operatorname{nm}%
%EndExpansion
^{2}\ $and a wire length $l=1%
%TCIMACRO{\unit{\U{3bc}m}}%
%BeginExpansion
\operatorname{\mu m}%
%EndExpansion
.$ We chose a damping of $\alpha=0.01$ and $P\beta_{c}^{\mathrm{mag}}=$
$P^{\prime}\beta_{Q}^{\mathrm{mag}}=P\gamma_{c}^{\mathrm{mag}}=P^{\prime
}\gamma_{Q}^{\mathrm{mag}}=1.$ The response function becomes dimensionless by
choosing appropriate units for the thermodynamic fluxes and forces. This leads
to%
\begin{equation}
\left(
\begin{array}
[c]{c}%
-\frac{\Delta V}{%
%TCIMACRO{\unit{mV}}%
%BeginExpansion
\operatorname{mV}%
%EndExpansion
}\\
-\frac{\Delta T}{0.1T}\\
\frac{\tau_{\mathrm{ext}}^{\mathrm{mech}}}{10^{-21}%
%TCIMACRO{\unit{N}}%
%BeginExpansion
\operatorname{N}%
%EndExpansion%
%TCIMACRO{\unit{m}}%
%BeginExpansion
\operatorname{m}%
%EndExpansion
}\\
-\frac{H_{\mathrm{ext}}}{0.1%
%TCIMACRO{\unit{T}}%
%BeginExpansion
\operatorname{T}%
%EndExpansion
}%
\end{array}
\right)  =\left(
\begin{array}
[c]{cccc}%
1.0 & 0.54 & -6.6\cdot10^{-4} & 0.66\\
0.054\, & 0.45 & 0 & 0\\
0.66 & 0 & 0.92 & 10^{4}\\
3.3\cdot10^{-3} & 0 & -0.05 & 0.5
\end{array}
\right)  \left(
\begin{array}
[c]{c}%
\frac{J_{c}}{%
%TCIMACRO{\unit{\U{3bc}A}}%
%BeginExpansion
\operatorname{\mu A}%
%EndExpansion
}\\
\frac{J_{Q}}{10^{-7}%
%TCIMACRO{\unit{J}}%
%BeginExpansion
\operatorname{J}%
%EndExpansion
/%
%TCIMACRO{\unit{s}}%
%BeginExpansion
\operatorname{s}%
%EndExpansion
}\\
\frac{\dot{\phi}}{%
%TCIMACRO{\unit{GHz}}%
%BeginExpansion
\operatorname{GHz}%
%EndExpansion
}\\
\frac{\dot{r}_{w}}{10^{5}%
%TCIMACRO{\unit{m}}%
%BeginExpansion
\operatorname{m}%
%EndExpansion
/%
%TCIMACRO{\unit{s}}%
%BeginExpansion
\operatorname{s}%
%EndExpansion
}%
\end{array}
\right)
\end{equation}
or%
\begin{equation}
\left(
\begin{array}
[c]{c}%
\frac{J_{c}}{%
%TCIMACRO{\unit{\U{3bc}A}}%
%BeginExpansion
\operatorname{\mu A}%
%EndExpansion
}\\
\frac{J_{Q}}{10^{-7}%
%TCIMACRO{\unit{J}}%
%BeginExpansion
\operatorname{J}%
%EndExpansion
/%
%TCIMACRO{\unit{s}}%
%BeginExpansion
\operatorname{s}%
%EndExpansion
}\\
\frac{\dot{\phi}}{%
%TCIMACRO{\unit{GHz}}%
%BeginExpansion
\operatorname{GHz}%
%EndExpansion
}\\
\frac{\dot{r}_{w}}{10^{2}%
%TCIMACRO{\unit{m}}%
%BeginExpansion
\operatorname{m}%
%EndExpansion
/%
%TCIMACRO{\unit{s}}%
%BeginExpansion
\operatorname{s}%
%EndExpansion
}%
\end{array}
\right)  =\left(
\begin{array}
[c]{cccc}%
1.1 & -1.3 & -7\cdot10^{-5} & -0.015\\
-0.13 & 2.4 & 2\cdot10^{-5} & 4\cdot10^{-3}\\
0.07 & -0.08 & 10^{-3} & -20\\
-8\cdot10^{-5} & 9\cdot10^{-5} & 10^{-4} & 2\cdot10^{-3}%
\end{array}
\right)  \left(
\begin{array}
[c]{c}%
-\frac{\Delta V}{%
%TCIMACRO{\unit{mV}}%
%BeginExpansion
\operatorname{mV}%
%EndExpansion
}\\
-\frac{\Delta T}{0.1T}\\
\frac{\tau_{\mathrm{ext}}^{\mathrm{mech}}}{10^{-21}%
%TCIMACRO{\unit{N}}%
%BeginExpansion
\operatorname{N}%
%EndExpansion%
%TCIMACRO{\unit{m}}%
%BeginExpansion
\operatorname{m}%
%EndExpansion
}\\
-\frac{H_{\mathrm{ext}}}{0.1%
%TCIMACRO{\unit{T}}%
%BeginExpansion
\operatorname{T}%
%EndExpansion
}%
\end{array}
\right)
\end{equation}

We can make a number of observations. For the present example the
self-consistency effects, \textit{e.g}. $L_{ii}^{-1}\neq\left(  \hat{L}%
^{-1}\right)  _{ii},\ $are well described by the perturbation approximation
used earlier, since the off-diagonal block matrices coupling of the
thermoelectric and magnetomechanical systems are rather small. These couplings
can be increased by a large diameter or shorter length of the wire. In
permalloy a temperature gradient of $0.2%
%TCIMACRO{\unit{K}}%
%BeginExpansion
\operatorname{K}%
%EndExpansion%
%TCIMACRO{\unit{nm}}%
%BeginExpansion
\operatorname{nm}%
%EndExpansion
^{-1}$ induces a charge current density of $10^{7}%
%TCIMACRO{\unit{A}}%
%BeginExpansion
\operatorname{A}%
%EndExpansion%
%TCIMACRO{\unit{cm}}%
%BeginExpansion
\operatorname{cm}%
%EndExpansion
^{-2},\ $which should suffice to move the domain wall in state-of-the-art
wires. A material with a smaller saturation magnetization and large
dissipative torques such as GaMnAs will be more susceptible to heat and charge
current-induced magnetization dynamics. The small friction of the
nanotube-lubricated rotation causes the strong coupling between the mechanical
degree of freedom and the magnetization dynamics. The best way to enhance the
coupled dynamics is the use of materials with a low Gilbert damping, however.

\section{Summary, extensions, and conclusions}

We derived the linear response matrix for a magnetic wire in contact with
electric and thermal reservoirs that can rotate along its axis.

Jen and Berger\cite{Jen2} observed domain wall motion in amorphous magnetic
alloys under a temperature gradient as small as 0.1 $%
%TCIMACRO{\unit{K}}%
%BeginExpansion
\operatorname{K}%
%EndExpansion
/%
%TCIMACRO{\unit{\U{3bc}m}}%
%BeginExpansion
\operatorname{\mu m}%
%EndExpansion
$ from the hot to the cold side. They offer two alternative explanations,
\textit{viz}. an entropic driving force in a domain wall gas,\cite{Berger85}
or a domain wall drag by the eddy currents induced by the anomalous Nernst
effect.\cite{Jen1} In thin wires such as addressed here both mechanisms are
unlikely to compete with the thermal spin-transfer torque.\cite{Hatami} The
domain wall displacement due to temperature dependence of magnetic
anisotropies as utilized by Miyakoshi \textit{et al.}\cite{Miyakoshi} should
not play a role in soft magnets such as permalloy, but temperature dependence
of pinning potentials can affect the dynamics, in principle.

Sinitsyn \textit{et al.}\cite{Sinitsyn} predicted a translational domain wall
motion under rotation of the magnetization texture, finding an identical
dependence of domain wall velocity with rotation frequency as we
do.\cite{Bretzel} However, Sinitsyn \textit{et al.}\cite{Sinitsyn} consider
damping in the laboratory frame of reference and not in the rotating frame.
Their predictions therefore hold for domain walls rotated relative to the
lattice with direct magnetic dissipation into the environment, whereas we
focus here on combined rotations of lattice and magnetization with mechanical friction.

We conclude that a moving domain wall pumps heat, which we might call domain
wall Peltier effect. A sizable cooling power may be associated with
magnetic-field induced domain wall motion. The domain wall drag by the thermal
spin transfer as well as the domain wall cooling can be computed
microscopically by the methods used by Hals \textit{et al}.\cite{Hals,Hals3}
and Starikov \textit{et al}.\cite{Starikov} Kovalev \textit{et al}%
.\cite{Kovalev09} independently obtained results for the interaction between
heat currents and magnetization. Proceeding from arbitrary one-dimensional
textures they illustrate their results by a spin-spiral model rather than a
single domain wall, however. Spin spirals should be sensitive to pinning
effects due to (near) commensurability with the underlying lattice and at the
wire terminals that are prone to suppress magnetization motion. For the heat
and charge current driven magnetomechanical motion this can actually be an
advantage. Kovalev \textit{et al}.\cite{Kovalev10} address thermal coolers by
moving magnetization textures, but found low efficiencies for conventional
magnetic materials. Hals \textit{et al}. also point out that in GaMnAs the
heating due to dissipation takes over any cooling effects already at moderate
domain wall velocities.\cite{Hals3}

The set-up in Figure 1 generates a charge current-induced mechanical torque by
domain wall motion, which is quite different from the mechanical torque that
is generated by a decaying spin accumulation,\cite{Zolfagharkhani} or the
spin-torque electromotor.\cite{Kovalev07}

The spin-torque motors based on moving domain walls have a drawback: they can
operate only with a single stroke, limited by the wire length over which the
domain wall can propagate. A similar problem has been encountered for the DC
electromotor, which has been solved by Faraday in the form of a commutator
that periodically inverts the sign of the mechanical torque. However, a pinned
texture (domain wall or spin spiral) as a rotor material solves this issue.
Such a material would not profit from the enhanced out-of-plane dissipative
torques predicted by Hals \textit{et al}.\cite{Hals}

Many protein-based molecular motors in the cell may be Brownian
motors\cite{Astumian} such as Feynman's ratchet and pawl,\cite{Feynman} in
which stochastic motion in the presence of a temperature or chemical potential
difference produces useful work. The present contraption also produces work
out of a temperature difference on the nanoscale, thus can be interpreted as a
realization of Feynman's ratchet, in which directionality provided by the
\textquotedblleft pawl\textquotedblright\ is replaced by the chirality of the ferromagnet.

The present scheme can be extended into different directions. An extension
from one to two-dimensional textures is necessary to treat vortex domain walls
in wider wires.\cite{Yang} The formalism is easily extended to describe the
coupled motion of charges, lattice, energy and spins as a function of harmonic
driving forces in the linear response regime. This would allow handling
torsional vibrations that can be used to observe the basic phenomena more
easily than a rotation.\cite{Bretzel} When normal metal contacts are attached,
spin currents and accumulations become explicit thermodynamic
variables.\cite{Hals2} The spin-Seebeck effect\cite{Uchida} and its Onsager
equivalent, the spin-Peltier effect, can then be handled. The Onsager
relations in many-terminal structures such as those used in studies of the
spin and anomalous Hall effects,\cite{Hanc} will be extended to the thermal
counterparts, such as the spin and anomalous Nernst, Ettingshausen, and
Righi-Le Duc effects.

In conclusion, we investigated the coupling of charge and heat currents with
magnetization and lattice for a realistic model system. All parameters can be
determined by independent experiments and are accessible to microscopic
calculations. On the basis of the response matrix we predict various magnetic
nanoscale heat engines and estimate the parameters that govern their efficiency.

\begin{acknowledgments}
We acknowledge helpful discussions with Alexey Kovalev and Kjetil Hals, who
pointed out mistakes in the first version of the manuscript. This work was
supported in part by the Dutch FOM Foundation, the Research Council of Norway,
Grants Nos.~158518/143 and 158547/431, and EC Contract IST-033749
\textquotedblleft DynaMax\textquotedblright, the Alfred P. Sloan Foundation,
and the NSF under Grant No. DMR-0840965. YT and GB would like to thank the
hospitality they enjoyed at the TU Delft and NTNU Trondheim, respectively.
\end{acknowledgments}


\begin{thebibliography}{99}
\bibitem {Note}In the limit of strong spin orbit scattering in which spin
currents and spin-dependent conductivities are not well defined,
Eqs.~(\ref{tauC}) and (\ref{tauQ}) still hold by intepreting the prefactors as
leading terms in a formal gradient expansion of charge and heat
currents.\cite{Kovalev09}
\end{thebibliography}
\end{document}